# Non-Hermitian global synchronization

*Weixuan Zhang*[#], *Fengxiao Di*[#], *and Xiangdong Zhang*[*]

Key Laboratory of Advanced Optoelectronic Quantum Architecture and Measurements of Ministry of Education, Beijing Key Laboratory of Nanophotonics and Ultrafine Optoelectronic Systems, School of Physics, Beijing Institute of Technology, Beijing 100081, China

[#]*These authors contributed equally to this work.*

[*]*Author to whom any correspondence should be addressed: zhangxd@bit.edu.cn*

**Abstract:** Synchronization of coupled nonlinear oscillators is a prevalent phenomenon in natural systems and can play important roles in various fields of modern science, such as laser arrays and electric networks. However, achieving robust global synchronization has always been a significant challenge due to its extreme susceptibility to initial conditions and structural perturbations. Here, we present a novel approach to achieve robust global synchronization by manipulating the interplay between non-Hermitian physics and nonlinear dynamics. Remarkably, the initial-state-independent non-Hermitian skin and topological global synchronization are proposed, exhibiting diverse anomalous effects such as the enlarged-size triggered non-Hermitian global synchronization and nonlinear skin states-dominated global synchronization. To validate our findings, we design and fabricate nonlinear topoelectrical circuits for experimental observation of non-Hermitian global synchronization. Our work opens up a promising avenue for establishing resilient global synchronization with potential applications in constructing high-radiance laser arrays and topologically synchronized networks.

## 1. Introduction

Synchronization, a collective oscillation behavior in which interacting units evolve in step with each other, has been extensively studied for over three centuries since Huygens' discovery of two coupled pendulum clocks with the same hunting frequency. Lately, synchronization has been observed and proven to play crucial roles in various fields such as electrical engineering, radio technology, biology, statistical mechanics and more. [1-8] To gain deeper insights into synchronization, extensive research on synchronized models such as the Kuramoto model and its derived models [9] have been conducted. These models have significant prospects in neuroscience, biochemistry, electronic information technology, and automation. It has been demonstrated that the pattern of network connections, coupling strengths and the frequency distribution of nonlinear oscillators can significantly affect the synchronized dynamics of nonlinear systems, rendering them susceptible to structural perturbations. Moreover, achieving synchronization in large-scale systems is highly contingent upon initial conditions. Thus, the accurate preparation of initial states is an indispensable prerequisite. These limitations pose significant challenges to the stability of synchronization. Therefore, the design of robust synchronized models of coupled nonlinear oscillators holds great significance in physics and technologies.

Over the past four decades, the study of topological insulators has attracted significant attention, [10-13] particularly due to the robust edge states within topological bandgaps, known as in-gap topological edge states, which offer a promising avenue for controlling nonlinear dynamics in a stable manner. [14-24] Recent theoretical investigations have further explored the interaction between these in-gap topological edge states and nonlinear synchronization, laying the groundwork for robust boundary synchronization [25, 26]. However, bulk oscillators in such systems often exhibit random or chaotic behavior, impeding the realization of global synchronization.

On the other hand, non-Hermitian Hamiltonians provide a versatile framework for a wide range of systems, from natural materials with intrinsic loss and gain to artificial structures with non-reciprocal couplings. [27-54] The interplay between synchronization and non-Hermitian physics has recently garnered increasing attention. For instance, a pioneering study demonstrated that non-reciprocal phase transitions can drive nonlinear synchronization, [27] where Kuramoto-like phase models and analytical mean-field diagrams were used to elucidate this exotic phenomenon. Furthermore, it has been demonstrated that phase synchronization can be achieved in a driven-dissipative system with bipartite non-Hermitian couplings between a single auxiliary mode and other oscillators. [28] While these studies highlight the pivotal role of non-Hermitian effects in synchronization, achieving global synchronization that is immune to initial conditions, scalable in size, and robust against structural perturbations remains an open challenge. One key feature of non-Hermitian systems is the non-Hermitian skin effect, where eigenstates localize at boundaries, with their number scaling with the system's volume. [30] Importantly, the non-Hermitian skin effect significantly breaks the orthogonality of eigenmodes, facilitating stronger mode coupling. This enhanced coupling enables modes to converge more readily into a global stable point or limit cycle, promoting robust synchronization that is largely insensitive to initial conditions across the system. Furthermore, the non-Hermitian skin effect can be leveraged to shape the distribution of topological in-gap states, [41] offering a novel mechanism to control the spatial profile of non-Hermitian synchronization dynamics. By harnessing these unique properties, non-Hermitian systems exhibiting the skin effect present a powerful platform for achieving robust, global synchronization, surpassing the limitations of traditional models.

In this work, we report the realization of non-Hermitian global synchronization that shows the strong robustness to initial conditions and structural perturbations. By tuning the coupling strength and lattice length, various types of non-Hermitian synchronization phenomena are proposed, including non-Hermitian linear skin-state synchronization, nonlinear skin-state synchronization, topological global synchronization, and skin-topological synchronized clusters. Leveraging the precise correspondence between electric circuit networks and tight-binding lattice models, we design and fabricate nonlinear topolectrical circuits to experimentally observe non-Hermitian global synchronization. Our findings provide a valuable framework for constructing stable global synchronization systems.

## 2. Non-Hermitian skin global synchronization.

We start to consider a Hatano-Nelson chain with onsite Stuart-Landau oscillators, as shown in **Figure 1a**. Non-reciprocal couplings are labeled by $J_{\pm}$ and Stuart-Landau oscillator is described by $\dot{Z}(t) = (i\omega_0 + \alpha - \beta|Z(t)|^2)Z(t)$ with the gain coefficient $\alpha$ and nonlinearity $\beta$. The right chart presents the self-excited oscillation of a Stuart-Landau oscillator with frequency and amplitude being $\omega_0$ and $\sqrt{\alpha/\beta}$. The lattice dynamical equation is

$$\dot{Z}_l = (i\omega_0 + \alpha - \beta|Z_l|^2)Z_l - i(J_+Z_{l+1} + J_-Z_{l-1}), \quad (1)$$

where $l$ is the site number from 1 to $N$. We calculate the wave evolution of the model with a random initial state $Z_l(t=0) \in [-0.1, 0.1]$, and other parameters are $N$=15, $\omega_0 = 0.1$, $\alpha = 5e^{-3}$, $\beta = 5e^{-4}$, $J_+ = 1.5$, $J_- = 1$. After a long period, the systematic dynamics invariably converge into one of two possible synchronized oscillations, as presented in **Figs. 1(b1)-(b2)**, where the real parts of wave amplitudes at odd and even lattice sites are plotted by black and red lines. Two insets present the locally enlarged views. It is worth noting that a thousand of random initial states have been tested, which all evolve into one of these two single-frequency oscillations. In addition, the phase differences between even and odd sites for these two synchronized dynamics are $\pi$ and 0, thus denoting them as anti-phase and in-phase synchronized states, respectively. We further perform the linear stability analysis by linearizing Eq. (1) around anti-phase and in-phase synchronized states, demonstrating that both synchronized states are stable with respect to perturbations (see Supporting Information 1). The appearance of these non-Hermitian global synchronization can be understood from the following three perspectives.

Firstly, when the dynamical condition $\beta|Z_l(t)|^2 \ll J_+,\ J_-$ is satisfied, the lattice waveform $|\boldsymbol{Z}(t)>=[Z_1(t),\dots,Z_N(t)]^T$ can be expanded by the linear eigenstates of the system with $\beta=\alpha=0$, corresponding to the Hatano-Nelson chain. To clarity, we express the nonlinear eigen-equation of our model as $-i\varepsilon\varphi_l=(i\omega_0+\alpha-\beta|\varphi_l|^2)\varphi_l-i(J_+\varphi_{l+1}+J_-\varphi_{l-1})$, where $\varepsilon$ represents the eigenenergy. $|\boldsymbol{\varphi}(\varepsilon)>=[\varphi_1,\dots,\varphi_N]^T$ is the eigenstate with superscript *T* short for transpose and $\varphi_l$ represents the amplitude at the *l*th site. Notably, $|\boldsymbol{\varphi}(\varepsilon)>=[\varphi_1,\dots,\varphi_N]^T$ is considered a linear eigenstate if its spatial profile closely matches the right eigenstate $|\boldsymbol{\varphi}^R(\varepsilon)>$ of the non-Hermitian linear eigen-equation with $\beta=\alpha=0$. Otherwise, it is referred to as a nonlinear eigenstate when the value of $\beta|\varphi_l|^2$ cannot be neglected compared to $J_+,\ J_-$. This occurs when either the lattice length or non-reciprocal coupling exceeds certain critical values (see followings). Therefore, when $\beta|Z_l(t)|^2 \ll J_+,\ J_-$, the dynamical waveform can be expanded in terms of the linear eigenstates as $|\boldsymbol{Z}(t)>=\sum_{n=[1,N]}C_n(t)|\boldsymbol{\varphi}^R(\varepsilon_n)>e^{-i(\varepsilon_n+\omega_0)t}$, where $C_n(t)$ is the coefficient of the *n*th linear right eigenstate $|\boldsymbol{\varphi}^R(\varepsilon_n)>$ with eigenenergy $\varepsilon_n$. Based on the bi-orthogonal relationship of right and left eigenstates of the linear non-Hermitian Hamiltonian, [38] we have $C_n(t)=<\boldsymbol{\varphi}^L(\varepsilon_n^*)|\boldsymbol{Z}(t)>$ with $<\boldsymbol{\varphi}^L(\varepsilon_n^*)|$ being the corresponding left eigenstate.

Secondly, the non-Hermitian skin effect significantly disrupts the orthogonality of linear eigenstates, as all eigenmodes become localized at one boundary. This results in strong coupling between initially excited non-Hermitian linear eigenstates during the wave evolution. It is worth noting that the strength of mode coupling depends on the system parameters, such as length *N* and non-Hermitian coupling coefficients. As the system evolves in the time-domain, the expansion coefficients $C_n(t)$ for different eigenstates change continually. In this case, the proposed non-Hermitian synchronization refers to the process in which only the coefficient of a linear eigenstate remains non-zero, while the coefficients of all other linear eigenstates decay to zero over time, corresponding to the transfer to the final synchronized state.

Finally, we explain why all initially excited eigenstates eventually converge to a single eigenmode. From Eq. (1), the system's effective Hamiltonian can be written as $H=\sum_l 0.5(-\alpha|Z_l|^2+0.5\beta|Z_l|^4)-i\sum_{l,k}J_{kl}Z_l^*Z_k$, with $J_{kl}=-\omega_0$ for $l=k$ and $J_{kl}=J_\pm$ for $l=k\pm1$. The dynamical equation follows from this effective Hamiltonian as $\dot{Z}_l=-\partial H/\partial Z_l^*$. Thus, the first term in the Hamiltonian represents the effective potential energy. In this framework, different eigenstates of the system experience different effective potentials due to the nonlinear term $\sum_l 0.25\beta|Z_l|^4$. The eigenstate that minimizes the potential energy corresponds to the state with the minimal inverse participation ratio (*IPR*), which is defined as $IPR(\varepsilon)=\sum_l|\boldsymbol{\varphi}_l(\varepsilon)|^4$. Due to the non-orthogonality of non-Hermitian eigenstates induced by the non-Hermitian skin effect, initially excited linear eigenmodes couple with one another, and the system is excepted to evolve into the linear eigenstate with the minimal effective potential—namely, the eigenstate with the minimal *IPR*, which minimizes the effective potential energy.

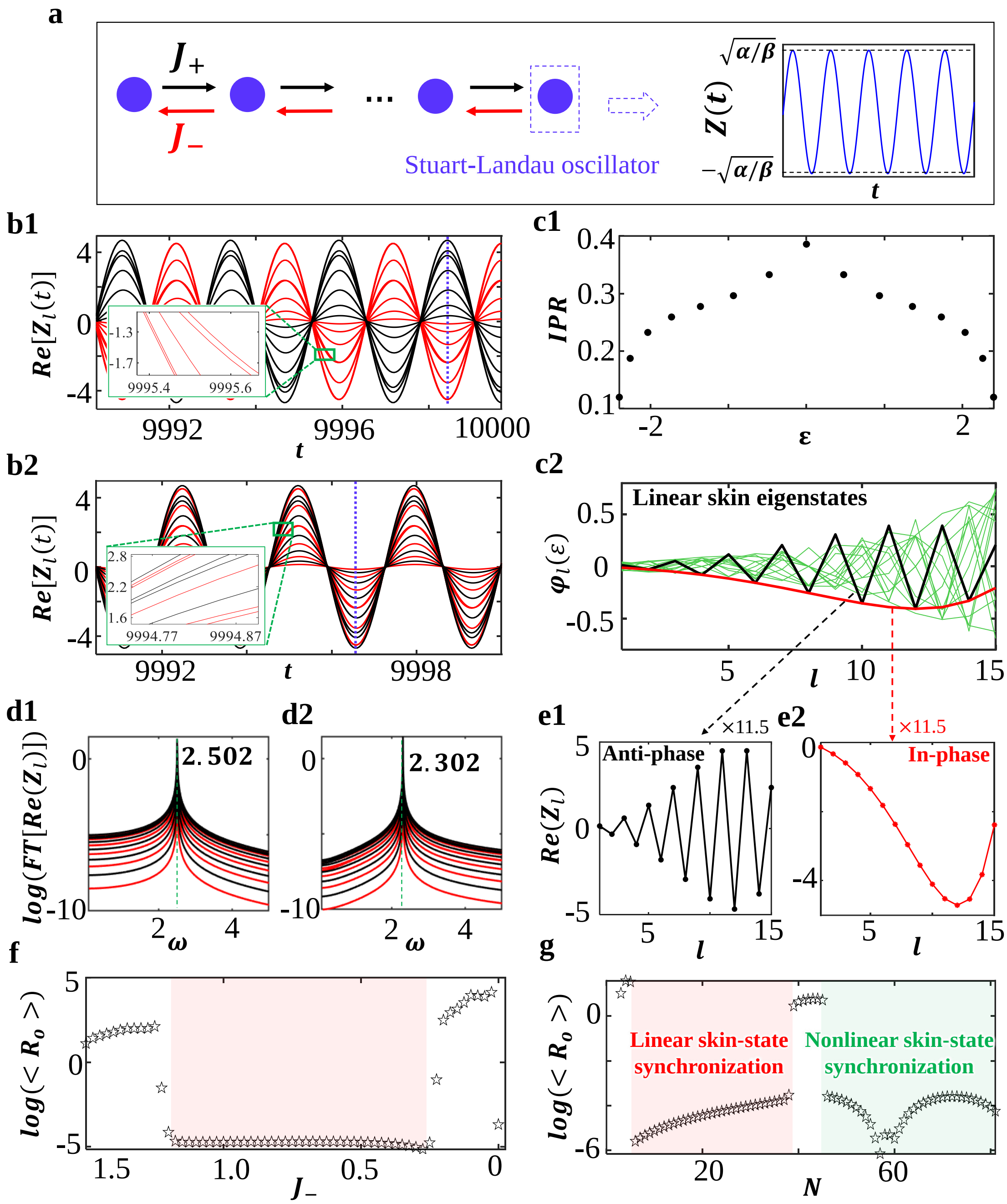

**Figure. 1. Theoretical results of non-Hermitian skin global synchronization.** (a). The scheme of the tight-binding lattice model for realizing non-Hermitian skin global synchronization. The Stuart-Landau oscillator is added at each lattice site, and the non-reciprocal coupling is used to couple nearest neighbored lattice sites. The right chart displays the wave evolution of a single Stuart-Landau oscillator. (b1) and (b2). Numerical

results of wave dynamics for the anti-phase and in-phase synchronized states. Black and red lines plot the real part of wave amplitudes at odd and even sites. Two insets present the locally enlarged views. It is noted that there are indeed 8 red lines, where two pairs of red lines are nearly overlapping. (c1). Numerical results of the linear eigen-spectrum accomplished by $IPR$ of each eigenstate. (c2). Spatial profiles of linear eigenstates for the non-reciprocal chain. Black and red lines correspond to linear eigenstates with minimum IPR. (d1)-(d2) and (e1)-(e2) Numerical results of frequency spectra and spatial profiles of anti-phase and in-phase synchronized states (at the time marked by blue lines in (b1)-b(2)). (f). The variation of the order-parameter $R_o$ as a function of the non-reciprocal coupling strength $J_-/J_+$ with $N = 15$. The red block marks the region sustaining non-Hermitian linear skin synchronization. (g). The numerical result of $R_o$ as a function of the lattice size $N$ with a fixed non-reciprocal strength ($J_+ = 1.5,\ J_- = 1$), where red and green blocks correspond to the regions of non-Hermitian linear skin-state synchronization and nonlinear skin-state synchronization, respectively. Here, the parameters related Stuart-Landau oscillator are set as $\omega_0 = 0.1,\ \alpha = 5e^{-3},\ \beta = 5e^{-4}$.

To numerically demonstrate above explanations, we calculate the eigenspectrum accompanied by $IPR$ of each eigenstate for the Hatano-Nelson chain, as presented in **Fig. 1(c1)**. Black and red lines in **Fig. 1(c2)** show spatial profiles of two minimum-IPR eigenmodes at $\varepsilon_1 = 2.402$ and $\varepsilon_2 = -2.402$. Green lines show other linear skin eigenstates. It is shown that all eigenstates are localized on the right boundary, manifesting the existence of non-Hermitian skin effect. Then, we perform discrete Fourier transform ($FT$) on two waveforms of **Figs. 1(b1) and 1(b2)**, as shown in **Figs. 1(d1) and (d2)**. We find that all lattice sites possess the identical oscillation frequency of $\omega_1 = 2.502$ or $\omega_2 = 2.302$, aligning with minimum-IPR eigenenergies of $\omega_1 = |\varepsilon_1 + \omega_0|$ and $\omega_2 = |\varepsilon_2 + \omega_0|$. The spatial profiles of anti-phase and in-phase synchronized states at the time marked by blue dashed lines in Fig. 1b are plotted in **Figs. 1(e1) and (e2)**, which are precisely matched to two minimal-IPR eigenstates presented by black and red lines in Fig. 1(c2) with the normalized factor being 11.5. These remarkable correspondences demonstrate the realization of synchronization facilitated by non-Hermitian linear skin states, which we refer to as the non-Hermitian linear skin-state synchronization.

To further analyze the influence of systematic parameters on the non-Hermitian skin global synchronization, an order-parameter should be used to distinguish synchronized and non-synchronized dynamics in our model. It is well known that the order-parameter defined as $R(t) = \left|\sum_{j=1}^{N} Z_l(t)\right|/N$ [9] is always used to characterize nonlinear synchronization. A high value of $R(t)$ indicates a phase-locked synchronized state, while it tends towards zero for non-synchronized states. While, *different* from conventional phase-locked synchronized states (correspond to the in-phase synchronized state in Fig. 1(b2)), our model exhibits both in-phase (depicted in Fig. 1(b2)) and anti-phase (depicted in Fig. 1(b1)) non-Hermitian skin synchronized states. Therefore, although the order-parameter $R(t)$ can effectively manifest in-phase non-Hermitian synchronized states, it fails to distinguish anti-phase non-Hermitian skin synchronized states (see Supporting Information 2 for details). Therefore, we apply an extended version of Kuramoto order parameter, denoted as $R_o = \{max[R(t)] - min[R(t)]\}_{t>t_0}$ ($t_0$ is sufficiently large to reach steady states) to quantify non-Hermitian global synchronization. In this case, we consistently find that the proposed order parameter approaches to zero for both in-phase and anti-phase non-Hermitian synchronized states, but exhibits larger values for non-synchronized states.

Using the order-parameter $R_o$, we firstly show that non-Hermitian skin synchronization depends on the non-reciprocal coupling strength. **Fig. 1f** presents the variation of $R_o$ as a function of $J_-$ with *N*=15 and $J_+ = 1.5$. Each point is averaged by thousands of initial states. We find that the non-Hermitian skin synchronization occurs only within the range of $0.25 \leq J_- \leq 1.2$ (marked in red). In contrast, non-synchronized states with large-valued $R_o$ emerge under the weak ($J_- > 1.2$) and strong ($J_- < 0.25$) non-reciprocal conditions. Supporting Information 3 presents the numerical results of the wave dynamics with $J_- = 1.4$ and $J_- = 0.05$. We find that the wave amplitudes possess small values ($|Z_l(t)| \sim 5$) with $J_- = 1.4 > 1.2$, thereby validating the effectiveness of linear-eigenstate expansion on the system dynamics with $\beta |Z_l|^2 \ll J_+,\ J_-$. Thus, the dynamical wave profile can be expressed as a superposition of all linear eigenstates, where the coupling strength between these eigenstates is determined by their orthogonality. With a fixed lattice length, the weaker non-reciprocal coupling (where the difference between $J_+$ and $J_-$ is small) results in a very weak non-Hermitian skin effect for the linear eigenstates, causing them to be nearly orthogonal to each other. The low degree of non-orthogonality results in an extremely weak coupling strength, preventing the initially excited eigenmodes from effectively transferring to the eigenmode with the minimal IPR. While, as for the case with $J_- = 0.05 < 0.25$, the site amplitudes are significantly enhanced ($|Z_l(t)| \sim 50$) with $\beta |Z_l|^2$ becoming comparable to the linear coupling terms. Consequently, the linear-eigenstate expansion of system's dynamics loses its effectiveness. Instead, the system's behavior is governed by the nonlinear eigenstates. However, due to the relatively weak non-orthogonality of these nonlinear eigenstates in this region, the system exhibits multi-frequency dynamics associated with various nonlinear eigenmodes.

Moreover, non-Hermitian skin global synchronization also exhibits a size-dependent critical behavior. **Fig. 1g** illustrates the variation of $R_o$ as a function of *N*. In the short-length region (*N*≤38), non-Hermitian linear skin-state synchronization occurs with *N*>7, showing an interesting phenomenon where larger systems exhibit a more coherent dynamical behavior. This is attributed to the weak non-orthogonality of non-Hermitian linear skin eigenstates with *N*≲7. As the lattice length increases, the non-orthogonality of these states is further enhanced, leading to the appearance of non-Hermitian linear skin-state synchronization. In the transition region ($39 \leqslant N \leqslant 45$), the order-parameter increases significantly, indicating the non-synchronized dynamics. To illustrate this, we calculate the dynamical evolution of the model with $N$ = 41 in Supporting Information 4. We observe that, in addition to the in-phase and anti-phase non-Hermitian skin synchronization, there is another dynamical behavior involving both in-phase and anti-phase non-Hermitian skin states. This phenomenon arises from the weak coupling between in-phase and anti-phase synchronized states within the transition region. To quantitatively assess the effective coupling between the in-phase and anti-phase synchronized states across different system sizes, we present their spatial profiles for various values of $N$ in Supporting Figure 7. We find that in the regions $7 < N \leq 38$ and $N > 45$, a larger proportion of lattice sites exhibit significant amplitudes for both in-phase and anti-phase skin synchronized states compared to the transition region ($39 \leq N \leq 45$). Thus, in the transition region, the limited number of large-amplitude lattice sites for both in-phase and anti-phase synchronized states leads to relatively weak effective coupling between them. This weaker coupling allows for the coexistence of both states during the system's dynamic evolution when these two states are simultaneously excited at the initial time. Beyond a critical length ($N$>45), the approximation of the system dynamics through linear-eigenstate expansion becomes ineffective. Because, as the system size increases, the linear skin modes become more localized, with many lattice sites displaying near-zero probability amplitudes. If we assume that the system's dynamics continue to follow the linear eigenstates, many oscillators on the side opposite the skin boundary experience the significant gain effect ($\alpha - \beta|Z_l|^2 \gg 0$). Stable synchronization requires a balance between gain and loss. To maintain self-oscillation, the amplitudes of oscillators near the skin boundary must increase substantially, reflecting the loss effects ($\alpha - \beta|Z_l|^2 \ll 0$). Consequently, the nonlinear term $\beta|Z_l(t)|^2$ can no longer be neglected, leading to a breakdown of the linear-eigenstate expansion. In this regime, the system ultimately synchronizes to the nonlinear eigenstate with the steady-state profile being matched to in-phase or anti-phase nonlinear eigenstate obtained by Newton-gradient method (See Supporting Information 5). The reappearance of non-Hermitian skin synchronization arises from the increased effective coupling length for in-phase and anti-phase nonlinear skin synchronized states. We refer to this phenomenon as the nonlinear skin-state synchronization.

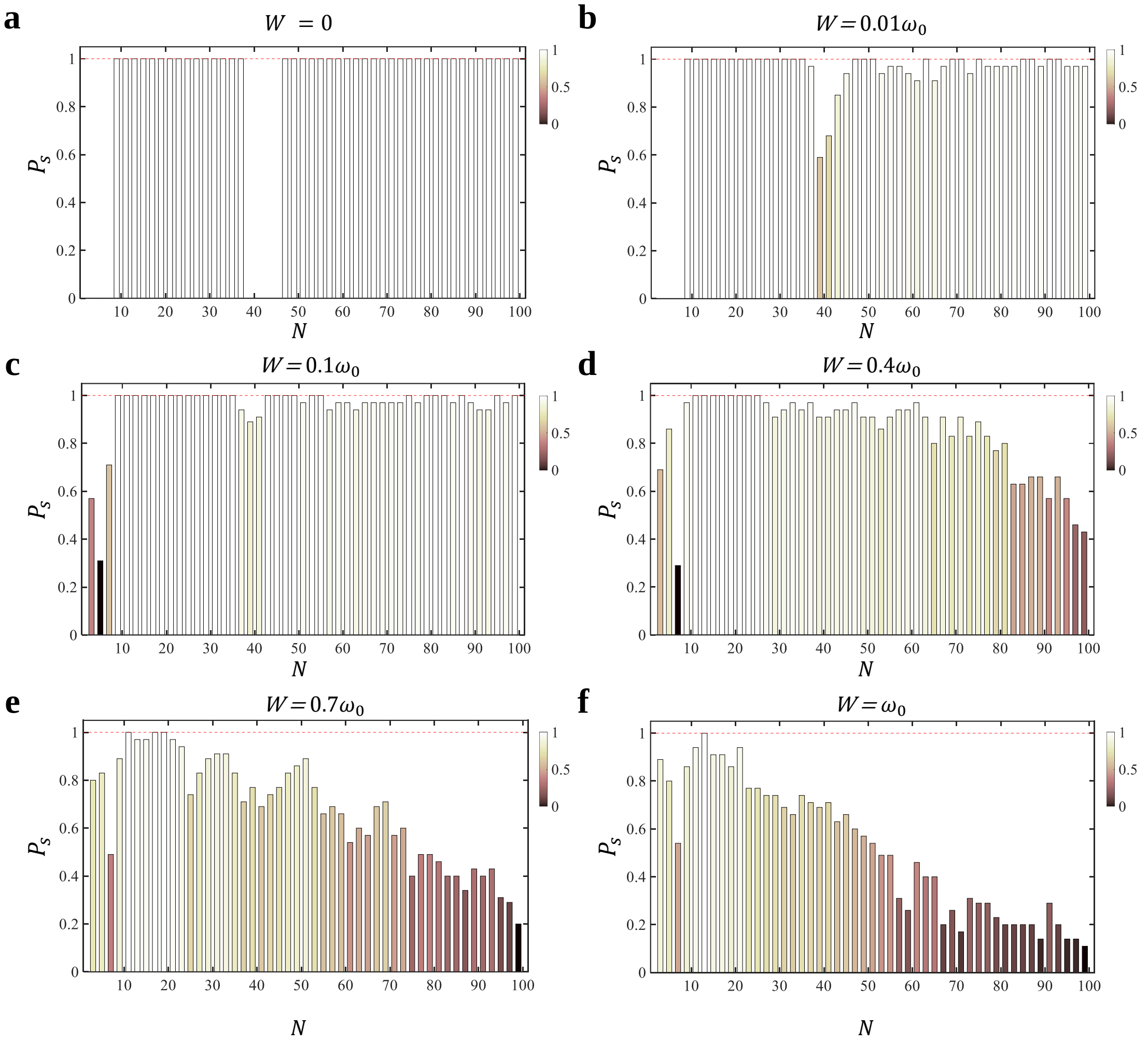

**Figure 2. The influence of disorder on non-Hermitian skin synchronization.** (a)-(f). The probabilities for the appearance of non-Hermitian global synchronization as a function of the system size with the disorder strength being $W = 0$, $0.01\omega_0$, $0.1\omega_0$, $0.4\omega_0$, $0.7\omega_0$ and $\omega_0$. Other parameters are identical to that used in Fig. 1g.

Lastly, we demonstrate that non-Hermitian skin global synchronization remains robust against weak structural perturbations. To illustrate this, we introduce random disorder to the self-oscillation frequency of each nonlinear oscillator, defined as $\omega_0 + [-W, W]$, where $W$ quantifies the strength of disorder. To avoid accidental results, we examine fifty random configurations with the same lattice length and disorder strength, and calculate the probability (labeled by $P_s$) of achieving the lattice model sustaining initial-state-immune non-Hermitian global synchronization. **Figures 2a-2f** present the variation of $P_s$ as a function of the system size with the disorder strength being $W = 0$, $0.01\omega_0, 0.1\omega_0, 0.4\omega_0, 0.7\omega_0$ and $\omega_0$, respectively. The color bar quantities the value of $P_s$. The results indicate that non-Hermitian skin synchronization persists with high probability under weak disorder. Notably, for disorder strengths being $W$=$0.01\omega_0$ and $0.1\omega_0$, the probability of achieving non-Hermitian global synchronization governed by linear eigenstates is 100% in the region of 7<$N$<39. Meanwhile, in the region governed by nonlinear eigenstates ($N$>45), the probability also exceeds 90%, showcasing the robustness of skin synchronization phenomena. Additionally, the previously unsynchronized regions ($N$<7 and 39<$N$<45) can also exhibit non-Hermitian global synchronization, albeit with certain probabilities. This occurs because disorder can alter the spatial profiles of eigenmodes and their effective couplings, allowing certain random configurations to support non-Hermitian global synchronization. As the disorder strength increases further, $P_s$ is significantly decreased in the entire length region, especially for larger system sizes. However, there remains a non-zero probability of observing non-Hermitian skin synchronization even in structures with strong disorder.

**3. Non-Hermitian topological global synchronization.**

Beyond the non-Hermitian skin synchronization, in this part, we construct non-Hermitian topological global synchronization with the linear coupling being the non-Hermitian SSH chain, as shown in **Figure 3(a)**. The last unit only contains the $A$-type sublattice. The reciprocal intercell ($J$) and non-reciprocal intracell ($J_\pm$) couplings are applied, along with a Stuart-Landau oscillator at each site. **Figs. 3(b)-(e)** plot spatial profiles of all linear eigenstates with $J$=0.2, 0.35, 0.5 and 1. Other parameters are $N$=25, $J_+ = 0.56$ and $J_- = 0.1$. Black and red (blue) lines correspond to minimal-IPR (other) bulk-band eigenmodes, which collapse towards the right end, revealing non-Hermitian skin effect. In addition, the bulk band of non-Hermitian SSH chain possesses nontrivial topology, [41] which induces topological edge states within the band gap. Green lines represent in-gap topological states at zero energy. It is noted that there is only one topological edge state. Because, the last unit of our model only contains the $A$-type sublattice. Eigenspectra and $IPRs$ with different $J$ are presented in Supporting Information 6. We find that the intercell coupling can manipulate the spatial profile of the in-gap topological state. It localizes on the same (opposite) boundary as skin states with $J$=0.2 and 0.35 ($J$=1). Whereas it extends throughout the entire bulk at $J$=0.5. In the following, we show that the spatial profile of the in-gap topological state plays a crucial role of non-Hermitian synchronization in the model. The dynamical equations are

$$\dot{Z}_{l,A} = (i\omega_0 + \alpha - \beta|Z_{l,A}|^2)Z_{l,A} - i(J_- Z_{l,B} + J Z_{l-1,B})$$
$$\dot{Z}_{l,B} = (i\omega_0 + \alpha - \beta|Z_{l,B}|^2)Z_{l,B} - i(J_+ Z_{l,A} + J Z_{l+1,A}). \quad (2)$$

Numerical results of the system dynamics with different $J$ are shown in **Figs. 3(f1)-(i1)** with $\omega_0 = 0.1$, $\alpha = 5e^{-4}$, $\beta = 5e^{-5}$.

The first system with *J*=0.2 exhibits the global synchronization, as shown in **Fig. 3(f1)**. The *FT* frequency spectrum in **Fig. 3(f2)** and steady-state profile in **Fig. 3(f3)** are aligned with the eigen-energy and spatial profile of a linear skin state with minimum localization (the red line in Fig. 3b). These results indicate that a linear skin state with minimum localization acts as global steady state, and the system undergoes non-Hermitian linear skin-state synchronization.

As for the second system with *J*=0.35, the wave amplitudes are significantly increased ($|Z_l(t)| \sim 50$), as illustrated in **Fig. 3(g1)**, making nonlinear eigenstates dominate the dynamical evolution. The steady-state frequency spectrum in **Fig. 3(g2)** and spatial profile in **Fig. 3(g3)** are both matched to the eigen-energy and eigen-profile of a nonlinear eigenstate calculated by Newton-gradient method (see Supporting Information 7), being nonlinear skin-state synchronization.

As for the third system with *J*=0.5, site amplitudes presented in **Fig. 3(h1)** are relatively small, manifesting the effectiveness on linear-eigenstate expansion of wave dynamics. We note that the linear topological state possesses the lowest localization with a minimum IPR compared to other skin modes (see Supporting Information 6). In addition, all linear eigenstates are localized at the same side, indicating that any pair of linear eigenstates possess the strong non-orthogonality. In this case, similar to the appearance of non-Hermitian linear skin-state synchronization, the systematic dynamics are ultimately governed by the in-gap topological states with the minimum IPR. The *FT* frequency spectrum in **Fig. 3(h2)** and synchronized profile in **Fig. 3(h3)** are both matched to the eigenenergy and spatial profile of the midgap topological zero mode (the green line in Fig. 3(d)). Such a correspondence shows the non-Hermitian global synchronization supported by midgap topological modes. We designate this phenomenon as non-Hermitian topological global synchronization that is robust with weak disorder (see Supporting Information 8).

In the last system with *J*=1, the small-amplitude waveform in **Fig. 3(i1)** suggests that linear eigenstates govern the evolution. Blue lines present the waveforms for eleven left-boundary sites, and red/black lines correspond to the right-boundary sites at odd/even positions. The *FT* frequency spectra and steady-state profile are shown in **Figs. 3(i2)-3(i3)**. Two peaks in the *FT* frequency spectra correspond to the eigen-energies of the in-gap topological state and the minimum-IPR linear skin state, respectively. Furthermore, the steady-state profile around the left (right) boundary shows a good correspondence to the midgap topological state (the minimum-IPR skin state), indicating that the midgap topological state (the minimum-IPR skin state) governs the local dynamics around the left (right) boundary. We refer to this phenomenon as the non-Hermitian skin-topological synchronized cluster. It arises because the midgap topological states and linear skin modes localize at opposite boundaries of the system. As a result, the dynamical evolutions of these two opposite boundaries are dominated by the topological edge state on one side and the skin states with the smallest IPR on the other. Due to their strong localization on opposite boundaries, these two states exhibit extremely weak coupling, preventing them from transitioning into a single synchronized state. This leads to the formation of the skin-topological synchronized cluster.

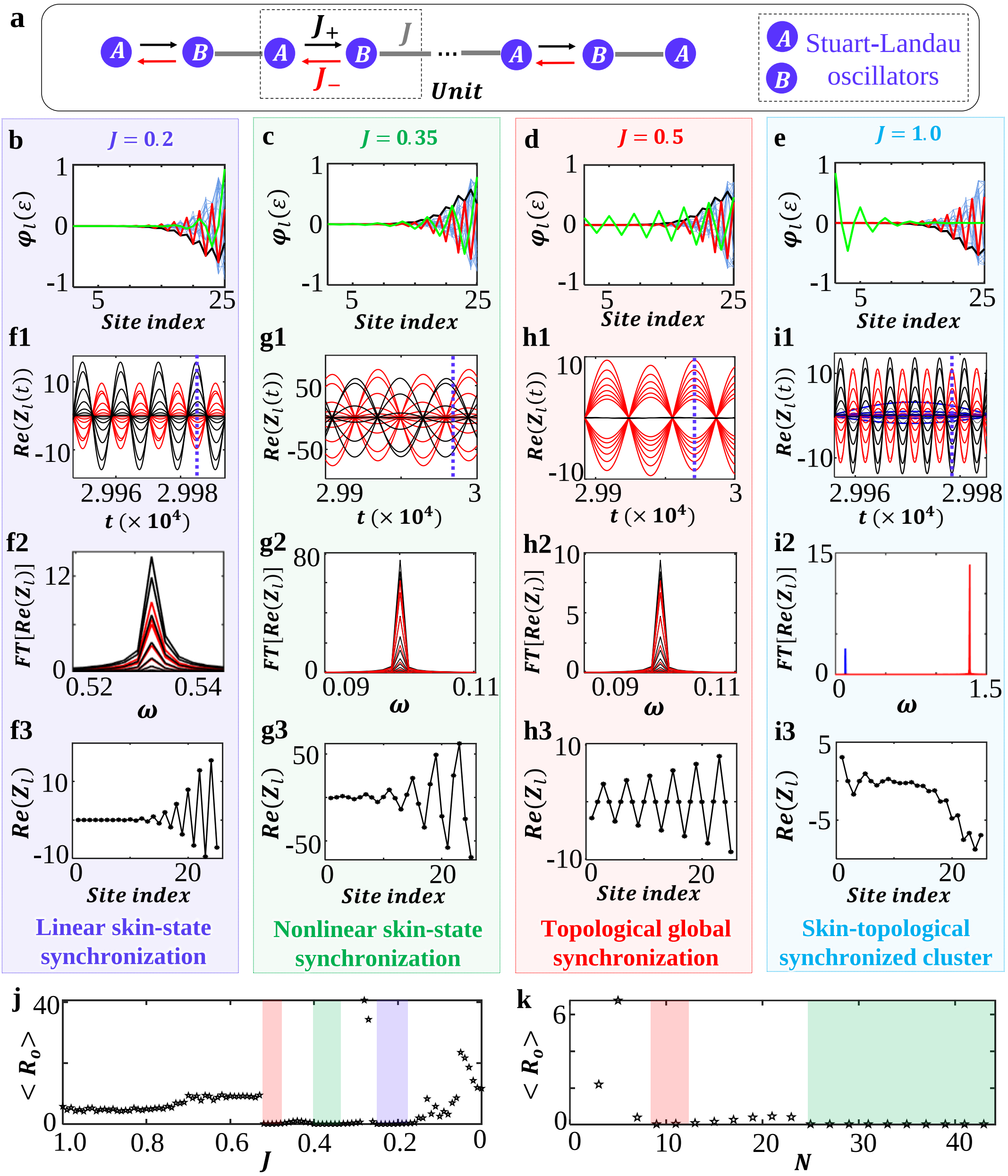

**Figure 3. Theoretical results of non-Hermitian topological global synchronization.** (a). The lattice model for the realization of non-Hermitian topological global synchronization. The reciprocal and non-reciprocal couplings are used for intercell and intracell couplings. The Stuart-Landau oscillators are added at each ‘A’ and ‘B’ sublattices. (b)-(e) Numerical results of the spatial profile of each linear eigenstate with J=0.2, 0.35, 0.5 and 1.0. (f1)-(i1). Numerical results of wave dynamics of all oscillators with the intercell coupling being $J = 0.2,\ 0.35,\ 0.5$ and $1.0$, respectively. The blue vertical dashed lines mark the time that we used to plot the spatial profiles in Figs. (f3)-(i3). (f2)-(i2) and (f3)-(i3) present the corresponding frequency spectra of all oscillators and steady-state spatial distributions of the system, respectively. (j). Numerical results for the variation of order parameter $R_o$ as a function of the intercell coupling strength $J$ with the lattice length being $N = 25$. (k). Numerical results for the variation of order parameter $R_o$ as a function of the lattice length $N$ with the intercell coupling strength being $J = 0.4$. Here, other parameters are set as $J_+ = 0.56$, $J_- = 0.1$, $\omega_0 = 0.1$, $\alpha = 5\mathrm{e}^{-4}$, and $\beta = 5\mathrm{e}^{-5}$. The appearances of the non-Hermitian linear skin-state synchronization, nonlinear skin-state synchronization and topological global synchronization are highlighted in purple, green and red blocks, respectively.

We further calculate the variation of order-parameter $R_o$ as a function of *J*, as shown in **Fig. 3j**. Three regions with the near-zero order-parameters appear, corresponding to three types of non-Hermitian global synchronization. Red, green, and purple domains correspond to non-Hermitian topological global synchronization, nonlinear skin-state synchronization, and linear skin-state synchronization, respectively. Transitions between different types of synchronization are related to the localization strength of the midgap topological state with respect to other linear skin states (See Supporting Information 6).

Additionally, the size-dependent critical behavior of non-Hermitian synchronization also exists in this model. The numerical result for the variation of $R_o$ as a function of the lattice length is shown in **Fig. 3k** with *J*=0.4. Two types of non-Hermitian global synchronization can appear in short- and long-length regions. In Supporting Information 9, the synchronized dynamics with *N*=11 and 31 are provided. We find that the site waveforms exhibit small amplitudes in the short-length region, and are dominated by the linear eigenstates of the non-Hermitian SSH chain. In this case, non-Hermitian topological global synchronization appears with 7<*N*≤13. While, in the long-length region, the system dynamics is governed by the nonlinear eigenstates. When the lattice length exceeds a critical value (*N*>23), nonlinear skin-state synchronization appears.

## 4. Experimental observation of non-Hermitian global synchronization.

Motivated by recent breakthroughs in realizing tight-binding lattice models using circuit networks, [55-73] in this part, we firstly design and fabricate nonlinear topoelectrical circuits to achieve non-Hermitian skin global synchronization. **Figure 4a** illustrates the RC circuit for realizing a pair of non-reciprocally coupled Stuart-Landau oscillators. It is important to note that the non-reciprocal coupling, implemented using a resistor-based impedance converter through current inversion (INIC), leads to a purely imaginary coupling term in the voltage dynamical equation, which does not match the real-valued non-reciprocal couplings $J_{\pm}$ in lattice models. To achieve real-valued coupling in the voltage dynamical equation, each lattice site is formed by four circuit nodes connected by nonreciprocal resistances $\pm R_w$. In this configuration, we can construct two voltage pseudospins at each effective lattice site, $V_{\uparrow i,\downarrow i} = V_{i,1} + V_{i,2}e^{\pm i\pi/2} + V_{i,3}e^{i\pi} + V_{i,4}e^{\pm i3\pi/2}$, which exhibit real-valued effective non-reciprocal couplings when two groups of adjacent circuit nodes are cross-connected by resistance $R_2$ in parallel with $\pm R_1$. Additionally, a Chua diode, which utilizes two multipliers and an INIC, is grounded at each node to serve as a Stuart-Landau oscillator. Each node also includes a grounded capacitor $C$ and a nonreciprocal resistance $\pm R_a$. In this setup, the effective nonlinear parameters are given by $\omega_0 = 2{R_w}^{-1}/C$, $\alpha = {R_3}^{-1}/C$ and $\beta = \frac{R_4+R_5}{100*R_3*R_4}/C$. The detailed correspondence between the Chua diode and Stuart-Landau oscillator is provided in Supporting Information 10.

By applying the Kirchhoff's law, the dynamical equation for two voltage pseudospins are expressed as (see Supporting Information 11 for the detailed derivation)

$$\frac{d}{dt}V_{i,\uparrow} = \left(j\frac{2}{CR_w}V_{i,\uparrow} + \frac{1}{CR_3} - \frac{R_4+R_5}{100CR_3R_4}\left|V_{i,\uparrow}\right|^2\right)V_{i,\uparrow} - j[\left(\frac{1}{CR_1} - \frac{1}{CR_2}\right)V_{i+1,\uparrow} + \left(\frac{1}{CR_1} + \frac{1}{CR_2}\right)V_{i-1,\uparrow}], \quad (3)$$

$$\frac{d}{dt}V_{i,\downarrow} = \left(-j\frac{2}{CR_w}V_{i,\downarrow} + \frac{1}{CR_3} - \frac{R_4+R_5}{100CR_3R_4}\left|V_{i,\downarrow}\right|^2\right)V_{i,\downarrow} + j[\left(\frac{1}{CR_1} - \frac{1}{CR_2}\right)V_{i+1,\downarrow} + \left(\frac{1}{CR_1} + \frac{1}{CR_2}\right)V_{i-1,\downarrow}], \quad (4)$$

where the circuit parameters are set to satisfy $\frac{1}{CR_a} - \frac{2}{CR_2} = 0$. We note that these equations share the same form as Eq. (1). Specifically, the effective tight-binding parameters are defined as $\omega_0 = \frac{2}{CR_w}$, $\alpha = \frac{1}{CR_3}$, $\beta = \frac{R_4+R_5}{100CR_3R_4}$, $J_+ = \frac{1}{CR_1} - \frac{1}{CR_2}$ and $J_- = \frac{1}{CR_1} + \frac{1}{CR_2}$ for Eq. (3), and $\omega_0 = -\frac{2}{CR_w}$, $\alpha = \frac{1}{CR_3}$, $\beta = \frac{R_4+R_5}{100CR_3R_4}$, $J_+ = -(\frac{1}{CR_1} - \frac{1}{CR_2})$ and $J_- = -(\frac{1}{CR_1} + \frac{1}{CR_2})$ for Eq. (4). Based on the mathematical correspondence between the dynamical equation of voltage pseudospins with the time-domain Schrödinger equation of the nonlinear non-Hermitian lattice model, our designed electric circuits can effectively simulate the non-Hermitian skin global synchronization.

An enlarged view of the fabricated circuit sample is shown in **Fig. 4b**, along with a photo of the nonlinear Chua diode. Circuit parameters are set as $R_w = 20k\Omega$, $R_1 = 0.8k\Omega$, $R_2 = 4k\Omega$, $C = 100nF$, $R_3 = 20k\Omega$, $R_4 = 10k\Omega$, and $R_5 = 90k\Omega$, which are sufficiently large to minimize the influence of parasitic effects in the circuit sample. We conduct the time-domain measurement of voltage dynamics in the fabricated circuit with *N*=5. Initial voltages of four circuit nodes in the first unit are set as [$V_{1,1}$, $V_{1,2}$, $V_{1,3}$, $V_{1,4}$]=[1, 0, -1, 0], while all other units have zero initial voltages.

As a result, both voltage pseudospins $V_{\uparrow 1}$ and $V_{\downarrow 1}$ are simultaneously excited. Notably, the two voltage pseudospins exhibit identical evolution when the system's dynamics can be approximated by a linear eigenstate expansion. This occurs because the eigenspectrum and eigenstates of the Hatano-Nelson chain remain unchanged when both the signs of $J_{\pm}$ and $\omega_0$ are simultaneously reversed. As a result, the evolution of voltages $V_{i,1}(t)$ and $V_{i,3}(t)$ under the simultaneous excitation of both pseudospins mirrors the dynamics observed when either $V_{\uparrow i}$ or $V_{\downarrow i}$ is excited individually. In this case, the voltage $V_{i,1}(t)$ behaves analogously to the wave amplitude $Z_i(\mathrm{t})$. **Fig. 4c** shows the measured voltage signals $V_{i,1}(t)$ at all sites (red and blue lines correspond to results at odd and even sites). The corresponding *FT* frequency spectra and steady-state voltage profiles are presented in **Figs. 4d and 4e**, respectively. It is shown that the voltage signal of each circuit oscillator exhibits the multi-frequency oscillation, and the long-time voltage profile corresponds to a superposition of multiple linear eigenmodes. These experimental results are consistent with simulations shown in Supporting Information 12, indicating that the short-length non-Hermitian circuit with *N*=5 do not exhibit the non-Hermitian skin global synchronization.

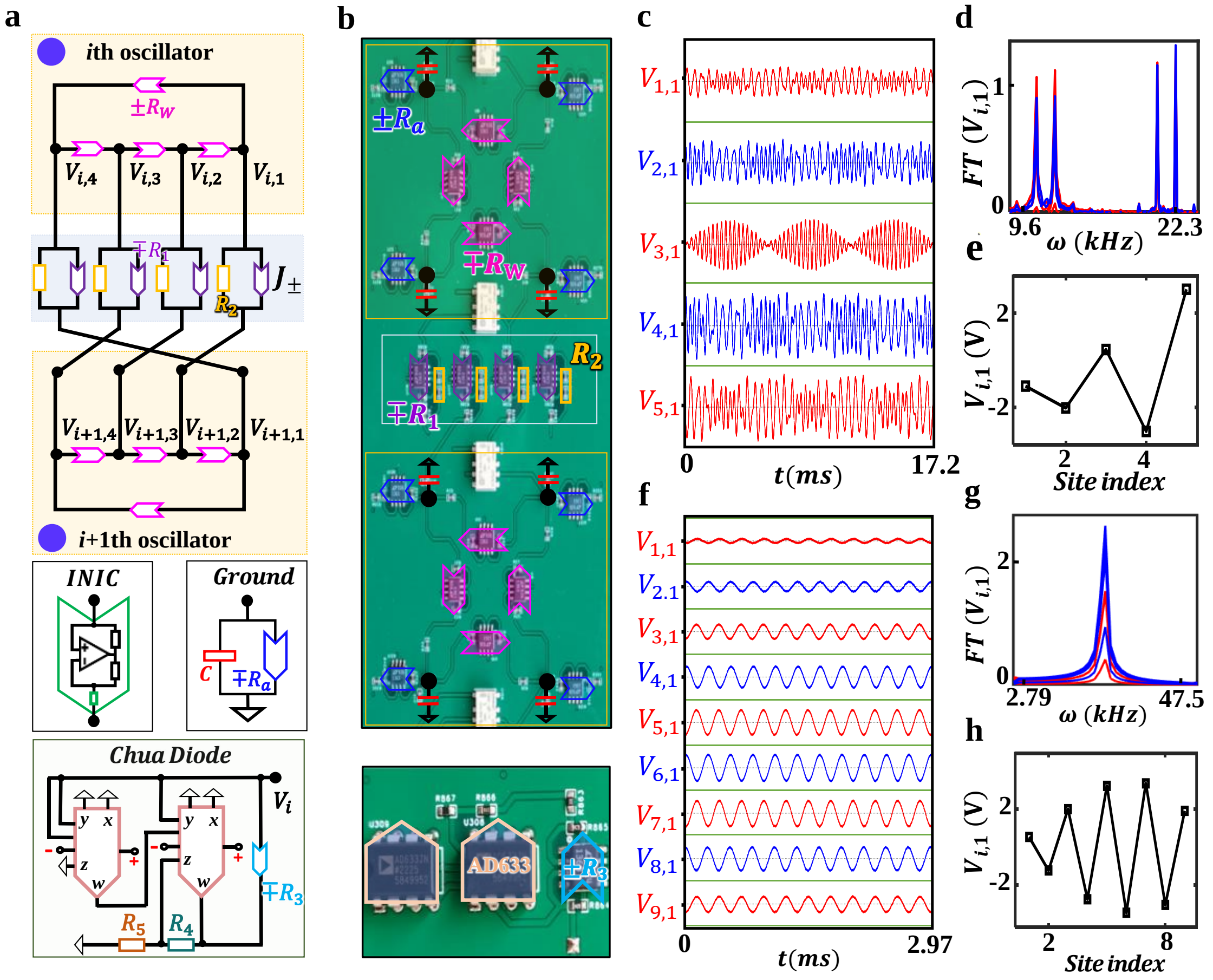

**Figure 4. Experimental results of non-Hermitian linear skin-state synchronization in electric circuits.** (a). The schematic diagram of the designed electric circuit to simulate the non-Hermitian linear skin-state synchronization. An effective lattice site is formed by four circuit nodes connected by nonreciprocal resistances $\pm R_w$, with each node grounded by a nonlinear Chua diode to act as the Stuart-Landau oscillator. Non-reciprocal coupling is achieved by crossly connecting adjacent circuit nodes using non-reciprocal resistance $R_2 \pm R_1$ to realize the non-reciprocal coupling. (b). A photograph image of two-coupled sites in the fabricated circuit sample for simulating non-Hermitian skin synchronization is shown, along with a bottom chart plotting the photo of a nonlinear Chua diode. (c)-(e). Measured voltage signals, the $FT$ frequency spectra and the steady-state voltage profile of the circuit sample with *N*=5. The multi-frequency dynamical evolution is observed. (f)-(h). Experimental results of the voltage evolution, the $FT$ frequency spectra and the steady-state voltage profile in the circuit sample with *N*=9. The non-Hermitian linear skin-state synchronization is realized. Other circuit parameters are set as $R_w = 20k\Omega$, $R_1 = 0.8k\Omega$, $R_2 = 4k\Omega$, $C = 100nF$, $R_3 = 20k\Omega$, $R_4 = 10k\Omega$, and $R_5 = 90k\Omega$.

Then, the voltage signals are measured in the other circuit sample with an extended length of *N*=9, as shown in **Fig. 4f**. The corresponding $FT$ frequency spectrum and steady-state voltage profile are illustrated in **Figs. 4g and 4h**. We find that all circuit oscillators exhibit the synchronized behavior with the oscillation frequency of 23.5 *kHz*, being consistent with the simulation results in Supporting Information 12. Furthermore, the steady-state voltage profile aligns with the non-Hermitian linear skin mode characterized by the minimal IPR. These experimental findings clearly demonstrate the achievement of non-Hermitian linear skin-state synchronization in electric circuits.

In addition to achieving non-Hermitian skin global synchronization, we further design nonlinear topoelectrical circuits to simulate the non-Hermitian topological global synchronization. The schematic of the circuit design for an effective unit cell with two sublattice sites is shown in **Figure 5a**. The realization of the non-reciprocal coupling follows the same approach as in Fig. 4a. The reciprocal inter-cell coupling is achieved by cross-connecting two groups of nodes (the different connection pattern compared to Fig. 4a) using non-reciprocal resistances $\pm R$, with the effective coupling strength being $J = 1/CR$. Additionally, each circuit node is grounded with a Chua diode (identical to the one used in Fig. 4), and different resistances $R_a$ and $R_b$ are employed for grounding on the '*A*' and '*B*' sublattices, respectively, to ensure the uniform effective dissipation across all nodes. In this setup, the voltage dynamical equation mirrors the form of Eq. (2) (see Supporting Information 11 for details).

**Fig. 5b** presents the image of a fabricated circuit unit with the following parameters $R_w = 20k\Omega$, $R_1 = 3.03k\Omega$, $R_2 = 4.35k\Omega$, $R = 2k\Omega$, $C = 1nF$, $R_3 = 200k\Omega$, $R_4 = 70k\Omega$, and $R_5 = 630k\Omega$. **Fig. 5c** plots the measured voltage signals $V_{i,a,1}(t)$ and $V_{i,b,1}(t)$ of all circuit oscillators with *N*=9. Here, Initial voltages are set as [$V_{1,a,1}$, $V_{1,a,2}$, $V_{1,a,3}$, $V_{1,a,4}$]=[1, 0, -1, 0], while all other nodes have zero initial voltages. The corresponding *FT* frequency spectra and steady-state voltage profile are shown in **Figs. 5d-5e**. It is shown that all circuit oscillators exhibit the multi-frequency oscillations, and the steady-state profile does not match to any linear eigenstate. To achieve non-Hermitian topological global synchronization, we follow the theoretical prediction in Fig. 3k, which suggests that increasing the lattice size can trigger the onset of non-Hermitian topological global synchronization. Thus, we fabricate and characterize another circuit sample with a larger size of $N$=15. The measured voltage signals are shown in **Fig. 5f**, with the corresponding *FT* frequency spectra and steady-state voltage profile presented in **Figs. 5g and 5h**, respectively. In this larger circuit, we observe that all oscillators synchronize at a frequency of 92.6 kHz, which corresponds to the eigenenergy of the midgap topological mode in the mapped lattice model. Additionally, the steady-state voltage profile aligns with the spatial distribution of the zero-energy midgap topological state, confirming the successful realization of non-Hermitian topological global synchronization. These experimental results are in excellent agreement with the simulations (see Supporting Information 13).

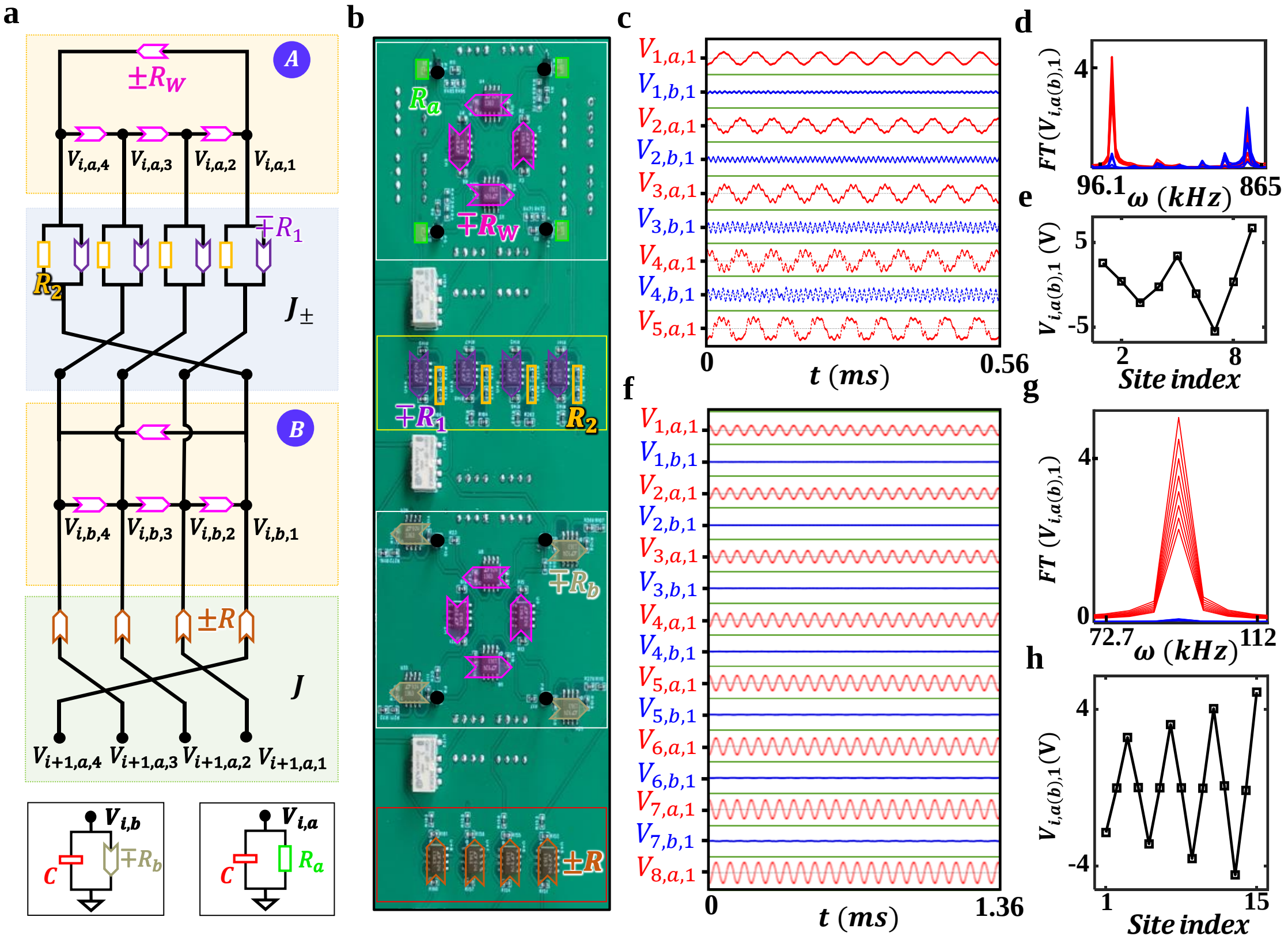

**Figure 5. Experimental results of non-Hermitian topological global synchronization in electric circuits.** (a). The schematic diagram illustrates the implementation of a single unit cell in an electric

circuit, featuring two sublattices, non-reciprocal intercell coupling, and reciprocal intracell coupling. Two bottom insets present the grounding of two types of sublattices. This setup is designed to simulate non-Hermitian topological global synchronization. (b). The photograph image of a single unit cell in the fabricated circuit, corresponding to the circuit diagram in (a). (c) and (f). Measured voltage signals of all nodes in two circuit samples with $N = 9$ and $15$. (d) and (e). The experimental results of $FT$ frequency spectra and distribution of steady-state voltage profiles in the short-length circuit with $N = 9$. (g) and (h) The measured $FT$ frequency spectra and steady-state voltage profile in the long-length circuit with $N = 15$. It is observed that non-Hermitian topological global synchronization exists in the circuit sample with a larger size. Here, other parameters are set as $R_w = 20k\Omega$, $R_1 = 3.03k\Omega$, $R_2 = 4.35k\Omega$, $R = 2k\Omega$, $C = 1nF$, $R_3 = 200k\Omega$, $R_4 = 70k\Omega$, and $R_5 = 630k\Omega$.

## 5. Discussion and conclusion

In conclusion, we have reported the first theoretical design and experimental realization of non-Hermitian skin and topological global synchronization. Compared to previous topological synchronization models, which are limited to the boundary synchronization effects with bulk oscillators exhibiting random or chaotic oscillations, our model firstly demonstrates the realization of non-Hermitian topological global synchronization. Furthermore, in contrast to traditional non-Hermitian synchronization models, our approach offers several advantages: it is immune to initial conditions, robust against structural perturbations, tunable in mode behavior, and scalable to larger systems. These features make it highly promising for a wide range of potential applications. In experiments, we designed and fabricated nonlinear topoelectrical circuits to observe non-Hermitian skin and topological global synchronization. Additionally, in Supporting Information 14, we further demonstrate the presence of non-Hermitian skin synchronization, where the non-Hermiticity is introduced solely through onsite loss and gain.

In practical applications, non-reciprocal couplings and onsite loss and gain are well-controlled non-Hermitian parameters in both classical and quantum systems, [27, 41-42, 48-52, 57] making our proposed non-Hermitian global synchronization feasible across a wide range of platforms. For instance, a recent study demonstrated a Hatano–Nelson laser array based on active optical oscillators, exhibiting both non-Hermiticity and nonlinearity, where phenomena such as the non-Hermitian skin effect, phase locking, and near-field beam steering were observed. [74] Our theoretical model can be applied to such photonic systems, potentially guiding new lasing behaviors driven by non-Hermitian global synchronization. Firstly, our non-Hermitian synchronized model can significantly enhance the stability of laser arrays against structural perturbations and pumping conditions. Moreover, our findings show that non-Hermitian global synchronization is scalable with engineered non-reciprocal couplings. Specifically, as the size of the Hatano–Nelson laser array increases, the non-reciprocal coupling can be tuned to proportionally scale the non-Hermitian skin mode, enabling non-Hermitian skin synchronization to persist in larger laser systems. This scalability is essential for practical applications, enabling the design of larger and more complex single-mode laser systems without compromising performance. Finally, by adjusting the non-reciprocal couplings, the spatial profile of non-Hermitian skin effects and topological synchronization can be finely tuned, allowing for a customizable near-field beam profile in the laser array. Additionally, beyond laser arrays, various artificial structures operating in the microwave region, such as split-ring resonators, [75] provide versatile platforms for investigating the interplay between non-Hermitian physics and nonlinear dynamics. In this context, achieving non-Hermitian global synchronization in the microwave domain holds great promise for applications in wireless power transfer and advanced sensing technologies.

**Acknowledgments.**

This work is supported by the National Key R & D Program of China No. 2022YFA1404900, National Science Foundation of China No. 12422411, Young Elite Scientists Sponsorship Program by CAST No. 2023QNRC001, Beijing Natural Science Foundation No. 1242027, and BIT Research and Innovation Promoting Project No.2023YCXZ020.

**Conflict of Interest**

The authors declare no conflict of interest.

**Data Availability Statement**

The data that support the findings of this study are available from the corresponding author upon reasonable request.

## References

[1] S. Strogatz, *Sync: The Emerging Science of Spontaneous Order. Physics Today*, **2003**.

[2] S.-B. Shim, M. Imboden, P. Mohanty, *Science* **2007**, 316, 95–99.

[3] J. Yan, M. Bloom, S. C. Bae, E. Luijten, S. Granick, *Nature* **2012**, 491, 578–581.

[4] A. Prindle et al., *Nature* **2012**, 481, 39–44.

[5] M. Rohden, A. Sorge, M. Timme, D. Witthaut, *Phys. Rev. Lett.* **2012**, 109, 064101.

[6] A. Jenkins, *Physics Reports*. **2013**, 525, 167-222.

[7] I. Nitsan, S. Drori, Y. E. Lewis, S. Cohen, S. Tzlil, *Nat. Phys.* **2016**, 12, 472-477.

[8] E. Post, M. C. Forchhammer, *Nature* **2002**, 420, 168-171.

[9] J. A. Acebrón, L. L. Bonilla, C. J. Pérez Vicente, F. Ritort, R. Spigler, *Rev. Mod. Phys.* **2005**, 77, 137-185.
[10] M. Z. Hasan, C. L. Kane, *Rev. Mod. Phys.* **2010**, 82, 3045-3067.
[11] X.-L. Qi, S.-C. Zhang, *Rev. Mod. Phys.* **2011**, 83, 1057-1110.
[12] T. Ozawa et al., *Rev. Mod. Phys.* **2019**, 91, 015006.
[13] G. Ma, M. Xiao, C. T. Chan, *Nat. Rev. Phys.* **2019**, 1, 281-294.
[14] D. Smirnova, D. Leykam, Y. Chong, Y. Kivshar, *Appl. Phys. Rev.* **2020**, 7.
[15] Y. Ota et al., Active topological photonics. *Nanophotonics* **2020**, 9, 547-567.
[16] T. Kotwal et al., *Proceedings of the National Academy of Sciences* **2021**,118, e2106411118.
[17] Z. Zhang et al., *Nat. Commun*, **2020**, 11, 1902.
[18] B. G.-g. Chen, N. Upadhyaya, V. Vitelli, *Proceedings of the National Academy of Sciences* **2014**,111, 13004-13009.
[19] D. Leykam, Y. D. Chong, *Phys. Rev. Lett.* **2016**, 117, 143901.
[20] Y. Lumer, Y. Plotnik, M. C. *Phys. Rev. Lett.* **2013**, 111, 243905.
[21] Y. Hadad, A. B. Khanikaev, A. Alù, *Phys. Rev. B* **2016**, 93, 155112.
[22] Y. Hadad, J. C. Soric, A. B. Khanikaev, A. Alù, *Nat. Electron.* **2018**, 1, 178-182.
[23] M. Jürgensen, S. Mukherjee, M. C. Rechtsman, *Nature* **2021**, 596, 63-67.
[24] M. Jürgensen, M. C. Rechtsman, *Phys. Rev. Lett.* **2022**, 128, 113901.
[25] C. W. Wächtler, G. Platero, *Phys. Rev. Res*. **2023**, 5, 023021.
[26] K. Sone, Y. Ashida, T. Sagawa, *Phys. Rev. Res*. **2022**, 4, 023211.
[27] M. Fruchart et al., *Nature* **2021**, 592, 363-369.
[28] J. Rohn, K. P. Schmidt, C. Genes, *Phys. Rev. A* **2023**, 108, 023721.
[29] C. M. Bender, S. Boettcher, *Phys. Rev. Lett.* **1998**, 80, 5243-5246.
[30] S. Yao, Z. Wang, *Phys. Rev. Lett.* **2018**, 121, 086803.
[31] K. Yokomizo, S. Murakami, *Phys. Rev. Lett.* **2019**, 123, 066404.
[32] K. Kawabata, K. Shiozaki, M. Ueda, M. Sato, *Phys. Rev. X* 2019, 9, 041015.
[33] S. Weidemann et al., *Science* **2020**, 368, 311-314.
[34] L. Xiao et al., *Nat. Phys.* **2020**, 16, 761-766.
[35] N. Hatano and D. R. Nelson, *Phys. Rev. Lett.* **1996**, 77, 570.
[36] L. Li, C. H. Lee, S. Mu, J. Gong, *Nat. Commun.* **2020**, 11, 5491.
[37] A. Ghatak, M. Brandenbourger, J. van Wezel, C. Coulais, *Proceedings of the National Academy of Sciences* **2020**, 117, 29561-29568.
[38] E. J. Bergholtz, J. C. Budich, F. K. Kunst, *Rev. Mod. Phys.* **2021**, 93, 015005.

[39] K. Zhang, Z. Yang, C. Fang, *Nat. Commun.* **2022**, 13, 2496.
[40] P. Gao, M. Willatzen, J. Christensen, *Phys. Rev. Lett.* **2020**, 125, 206402.
[41] W. Wang, X. Wang, G. Ma, *Nature* **2022**, 608, 50-55.
[42] K. Wang, A. Dutt, K. Yang, C. C. Wojcik, J. Vučković and S. Fan, *Science* **2021**, 371, 1240-1245.
[43] S. Longhi, *Annalen der Physik* **2018**, 530, 1800023.
[44] S. Wong, S. S. Oh, *Phys. Rev. Res.* **2021**, 3, 033042.
[45] T. Gao et al., *Nature* **2015**, 526, 554-558.
[46] L. Xiao et al., *Nat. Phys.* **2020**, 16, 761-766.
[47] S. Longhi, *Phys. Rev. Lett.* **2020**, 124, 066602.
[48] M.-A. Miri and A. Alù, *Science* **2019**, 363, eaar7709.
[49] L. Feng, R. El-Ganainy, and L. Ge, *Nat. Photonics.* **2017**, 11, 752.
[50] R. El-Ganainy, K. G. Makris, M. Khajavikhan, Z. H. Musslimani, S. Rotter, and D. N. Christodoulides, *Nat. Phys.* **2018**, 14, 11.
[51] R. Bouganne, M. Bosch Aguilera, A. Ghermaoui, J. Beugnon, and F. Gerbier, *Nat. Phys.* **2020**, 16, 21.
[52] N. Syassen, D. M. Bauer, M. Lettner, T. Volz, D. Dietze, J. J. García-Ripoll, J. I. Cirac, G. Rempe, and S. Dürr, *Science* **2008**, 320, 1329.
[53] A. McDonald, R. Hanai, A. A. Clerk, *Phys. Rev. B* **2022**, 105, 064302.
[54] S. Begg, R. Hanai, *Phys. Rev. Lett.* **2024**, 132, 120401.
[55] C. H. Lee et al., *Commun. Phys.* **2018**, 1, 39.
[56] J. Ningyuan, C. Owens, A. Sommer, D. Schuster, J. Simon, *Phys. Rev. X* **2015**, 5, 021031.
[57] V. V. Albert, L. I. Glazman, L. Jiang, *Phys. Rev. Lett.* **2015**, 114, 173902.
[58] S. Imhof et al., *Nat. Phys.* **2018**, 14, 925-929.
[59] T. Helbig et al., *Nat. Phys.* **2020**, 16, 747-750.
[60] N. A. Olekhno et al., *Nat. Commun.* **2020**, 11, 1436.
[61] P.-Y. Chen et al., *Nat. Electron.* **2018**, 1, 297-304.
[62] J. Wu et al., *Nat. Electron.* **2022**, 5, 635-642.
[63] Z. Wang, X.-T. Zeng, Y. Biao, Z. Yan, R. Yu, *Phys. Rev. Lett.* **2023**, 130, 057201.
[64] L. Song, H. Yang, Y. Cao, P. Yan, *Nat. Commun.* **2022**, 13, 5601.
[65] H. Yang, L. Song, Y. Cao, P. Yan, *Nano Lett.* **2022**, 22, 3125-3132.
[66] R. Li et al., *Natl. Sci. Rev.* **2020**, 8.
[67] S. Liu et al., *Research* **2021**.

[68] B. Lv et al., *Commun. Phys.* **2021**, 4, 108.

[69] Y. Yang, D. Zhu, Z. Hang, Y. Chong, *Sci. China Phys. Mech. Astron.* **2021**, 64, 257011.

[70] X. Zhang et al., *Commun. Phys.* **2023**, 6, 151.

[71] Y. Wang, L.-J. Lang, C. H. Lee, B. Zhang, Y. D. Chong, *Nat. Commun.* **2019**, 10, 1102.

[72] H. Hohmann et al., *Phys. Rev. Res*. **2023**, 5, L012041.

[73] M. Di Ventra, Y. V. Pershin, C.-C. Chien, *Phys. Rev. Lett.* **2022**, 128, 097701.

[74] Liu, Y.G.N., Wei, Y., Hemmatyar, O. et al. *Light Sci Appl* **2022**,11, 336.

[75] Z. Guo, Y. Wang, S. Ke, X. Su, J. Ren, H. Chen, *Adv. Phys. Res.* **2023**, 3, 2300125.